# Controllable High-Speed Rotation of Nanowires

D. L. Fan[1], F. Q. Zhu[2], R. C. Cammarata[1], and C. L. Chien[2*]

Department of Materials Science and Engineering[1]

Department of Physics and Astronomy[2]

Johns Hopkins University

Baltimore, MD, 21218

We report a versatile method for executing controllable high-speed rotation of nanowires by AC voltages applied to multiple electrodes. The rotation of the nanowires can be instantly switched on or off with precisely controlled rotation speed (to at least 1800 rpm), definite chirality, and total angle of rotation. We have determined the torque due to the fluidic drag force on nanowire of different lengths. We also demonstrate a micromotor using a rotating nanowires driving a dust particle into circular motion. This method has been used to rotate magnetic and nonmagnetic nanowires as well as carbon nanotubes.

---

[*] Author to whom correspondence should be address, email: clc@pha.jhu.edu

In the intense exploration of nanoscience and nanotechnology, it is often desirable, and indeed necessary, to manipulate small entities. The manipulation of small objects with sizes less than 10 μm encounters several formidable obstacles. Small objects are usually suspended in a suitable liquid to avoid the perils of adhering to the surface by the van der Waals forces. However, the motion of small entities in a liquid is in the realm of very low Reynolds numbers (defined as $\mathrm{Re} = \dfrac{D_p v \rho}{\eta}$, where $D_p$, $v$, $\rho$ and $\eta$ are size of particle, relative velocity, density of medium and viscous coefficient, respectively), where the drag force overwhelms the motion of the entities [1-3]. Consider a nanowire 10 μm in length and 0.3 μm in diameter suspended in deionized (DI) water, for which the Reynolds number is only about $10^{-5}$. Such a small Reynolds number dictates that, in the absence of other external forces, the drag force alone will stop a moving nanowire at 100 μm/sec in $10^{-6}$ sec within a short distance of 1 Å. By the same token, disproportionately large forces are required to overcome the drag force of a small entity and to initiate appreciable motion. Under a constant external force, the nanowire reaches the terminal velocity essentially instantly with no apparent acceleration.

It is even more challenging to controllably rotate nanoscale entities, especially with high rotation speeds. The problem is further compounded by the fact that the terminating torque due to the viscous drag force, that ultimately dictates the rotation of a small rotating entity in small Reynolds numbers, is not well known. These difficulties notwithstanding, a rotating magnetic field has been used to rotate ferromagnetic nanowires, such as Ni, albeit with low rotation rates [4, 5]. Optical tweezers using circularly polarized light and elaborate instrumentation can also rotate small particles, usually one particle at a time [6].

In contrast, the electric (*E*) field provided by patterned microelectrodes can accommodate force variations on the μm scale necessary for motion of small entities, and equally important, involves no moving parts. We have recently shown [7] that nanowires in suspension can be driven by an AC electric field to execute linear motion with high efficiency despite extremely low Reynolds number at the level of $10^{-5}$. We have determined the functional form of the force, and its dependence on the voltage and frequency of the AC field. In this work, we demonstrate a versatile method for executing controllable high-speed rotation in nanowires using a rotating electric field provided by AC voltages applied to multiple electrodes. The rotation of the nanowires can be instantly switched on or off with precisely controlled chirality, rotation speeds (to at least 1800 revolution per minute (rpm)), and total angle of rotation. This method also allows the determination of the viscous torque on nanowires of the same diameter but different lengths. We further show that this method can rotate various elongated metallic entities, including nonmagnetic and magnetic nanowires as well as multiwall carbon nanotubes (MWCNT). We have also demonstrated a micromotor using a bent Au nanowire in a rotating electric field, of relevance to small fluidic or mechanical devices, such as micro electromechanical systems (MEMS) [8, 9].

The gold (Au), platinum (Pt), and nickel (Ni) nanowires used in this work, 300 nm in width and lengths from 2 μm to 30 μm, were fabricated by electrodeposition through nanoporous alumina templates as described previously [7, 10]. After the alumina template was dissolved in a 4M NaOH solution, the nanowires were sonicated, centrifuged alternately in DI water and ethanol twice and then suspended in DI water with a low conductivity of 2.4 μSiemens/cm. The electrodes used to rotate the nanowires were made of Au patterned by laser micromachining on quartz substrates. Small amounts of 2-4 µl nanowires suspended in DI water were applied in the regions of Au electrodes as shown in Fig. 1a for executing rotational motion of the nanowires.

We used four separate electrodes with a gap distance between the opposite electrodes of about either 150 µm or 320 µm. Four AC voltages of the same magnitude and frequency but with a sequential phase shift of $90^0$ were simultaneously applied at the electrodes as shown in Fig. 1a. These voltages give rise to a rotating electric ($E$) field that compels all the suspended nanowires in the central region to rotate simultaneously. An optical microscope equipped with a video system with 30 frames per second captured the rotational motion of the nanowires. The angular speed ($\omega$) of an individual nanowire was determined by measuring the amount of rotation in fixed time intervals.

Most of measurements have been made using Au nanowires of the same length of 15 µm in quadruple electrodes with a 150 µm gap. Both free-standing nanowires and nanowires with one-end fixed by the thoil chemical linkage to the quartz substrate [11, 12] can be rotated. Under the same AC voltage with the same frequency, free-standing nanowires always rotate faster than those with one end fixed to the substrate as shown in Fig. 1b, where snapshots taken 1/30 sec intervals under $V_{AC}$ = 2.5 V at 80 kHz are overlapped.

The measured rotation speed $\omega$ of a nanowire depends both on the magnitude and the frequency of the applied AC voltage. The rotation rate increases quadratically with voltage as shown in Fig. 2a with slopes of 4.5 rpm/$V^2$ at 5 kHz, 18.1 rpm/$V^2$ at 80 kHz for free nanowires, and 6.3 rpm/$V^2$ at 80 kHz for nanowires with one-end fixed. The $V^2$ dependence is particularly advantageous for achieving high rotation speeds. Indeed, we have achieved very high rotation rates of 1803 rpm for free nanowires and 445 rpm for nanowires with one-end fixed at 10 V and f = 80 kHz as shown in Fig. 2a. We have detected no limitation in the rotation speed up to 1800 rpm, beyond which our video system with 30 frames per second could not accurately track the rotation. The value of ω also depends on the frequency of the AC voltage. As shown in Fig. 2b, at $V_{AC} = 2.5$ V and 5 V, the rotation rate of a free nanowire increases sharply below 50 kHz before leveling off and then decreasing slightly from 50 kHz to 300 kHz.

The ability to control the chirality of rotation for all the nanowires within the region of the electrodes is an important attribute. Since the phase shifted AC voltages applied to the quadruple electrodes set the rotation, the chirality of rotation is likewise determined by the phases of the AC voltages. Using the geometry shown in Fig. 1a, the rotation chirality is always the same as that of the field rotation but opposite to the phase shift direction of the AC voltages. The rotation is clockwise (or counter-clockwise) when the phase shift was $-90^0$ (or $90^0$). This fact has been verified in the frequency range of 5 kHz to 300 kHz.

To further illustrate the ability to control the rotation chirality, we repeatedly reversed the phase shifts of the applied AC *E* field. The nanowires responded to the reversed chirality and instantly reached the constant terminal rotation speeds as shown in Fig. 2c, with no apparent acceleration and deceleration detected within 1/30 sec. Furthermore, the rotation starts and stops as soon as the electric field was switched on and off. This apparent instant response is due to the extremely small Reynolds number of $10^{-5}$ for nanowires suspended in DI water. These results show that we can use the AC voltages to instantly switch on and off the rotation of nanowires, precisely control the rotation speed, the chirality of rotation, and the total angle of rotation.

The rotation of nanowires is due to the interaction [13] between the AC $E$ field and the induced electric dipoles of the nanowires. The charge carriers in a metallic entity are polarized by the electric field resulting in an electric dipole moment, whose value depends on the material and the geometrical shape of the object, and it is proportional to the value of the $E$ field. The induced dipole moment of a metallic sphere can be readily calculated. The dipole moment is substantially enhanced for an elongated ellipsoid, to which a metallic nanowire can be approximated. The large aspect ratio of the Au nanowires used here gives rise to an enhancement factor of about 380, which greatly increases the efficiency in rotating the nanowires [7]. The rotation of a nanowire is due to the torque $\boldsymbol{p} \, x \, \boldsymbol{E}$, where $\boldsymbol{p}$ is the induced dipole moment along the nanowire axis and $\boldsymbol{E}$ the rotating electric field, hence the torque varies as $E^2$. For a conducting metallic nanowire of length $L$ and radius $a$ suspended in a media of permittivity of $\varepsilon_m$, the magnitude of the torque is [14]

$$T_e \approx \frac{\pi L^3}{12} \frac{\varepsilon_m}{\ln\dfrac{L}{a}-1} E^2 , \qquad\qquad (1)$$

for $L >> a$. The $E^2$ dependence accounts for the observed $V^2$ dependence of the rotation rate as experimentally observed.

The actual rotation speed ω of the rotating nanowire is dictated by $T_e = T_\eta$, where $T_\eta$ is the torque due to the fluidic drag force. Because of the low Reynolds number, at each AC $E$ field, the rotating nanowires reaches a terminal angular velocity essentially instantaneously. Consequently, the rotation of the nanowire can be instantly switched on and off.  Also due to the large drag force, the nanowires cannot synchronize its rotation with that of the rotating field.  There is always a lag between the nanowire rotation and the $E$ field.  The chirality of rotation is determined by the charge relaxation times of the nanowire relative to that of the media in which the nanowires in embedded.  When the charge relaxation time of the nanowire is shorter than that of the media, as in the case of Au nanowire in DI water, the Au nanowires always rotate in the same chirality as that of the $E$ field as observed for the frequency range of 5 kHz to 300 kHz.

For a sphere of radius $R$, the torque $T_\eta$ has the form of $\eta\omega(8\pi R^3)$, where $\eta$ is the viscous coefficient, $\omega$ the rotation speed, and the last term is the geometrical factor for spheres.   For small entities of other shapes, including nanowires, the geometrical factor is not theoretically well known.  For nanowires of the same diameter, $T_\eta$ has the form of $\eta\omega f(L)$, in which $f(L)$ depends only on the length $L$.  Since the driving torque $T_e$ due to the AC field is known, we can in turn experimentally determine $f(L)$ by noting $T_\eta = \eta\omega f(L) = T_e \approx CL^3V^2$ from Eq.(1), where $C$ is a constant.  For this purpose, we have measured the values of $\omega$ of various Au nanowires of the same diameter of 300 nm with lengths ranging from 1.7 μm to 11 μm in quadruple electrodes with a 320 μm gap at 10 kHz applied voltages.  The measured dependence of $\omega/V^2$ on $L$ is shown in Fig. 3a.  The value of $\omega/V^2$ is weakly dependent on $L$ showing a maximum at $L=$ 6 μm.  To extract $f(L)$, we fit $\omega/V^2$ to $L^3/g(L)$, where the best-fit form is $g(L) = -0.00052L^6+0.0138L^5-0.11L^4+0.41L^3$, shown by the solid curve in Fig.3a.  We then plot $L^3V^2/\omega = g(L) =(\eta/C)f(L)$, which increases monotonically with $L$ as shown by the solid curve in Fig.3b.  In this manner, the elusive geometrical factor $f(L)$ for nanowires with 300 nm in diameter and length $L$ has been determined experimentally.  Note that within the length range, the $L^3$ term is the dominant one in $f(L)$ with the largest coefficient.  However, if $f(L)$ strictly includes only the $L^3$ term, then $\omega/V^2$ would have been independent of $L$.  Thus the variation of $\omega/V^2$ showing in Fig. 3a is a direct consequence of higher order terms in addition to the $L^3$ term in $f(L)$.

We have also used this versatile method to rotate other elongated entities including Au, Pt and Ni nanowires (300 nm in diameter and 9 μm in length), and multiwall carbon nanotubes (MWCNT) (50 nm in diameter and 5 μm in length) shown in Fig. 3a, with images taken every 1/15 sec, at 6 V, 10 kHz.  Using a quadruple electrode with 320 μm in separation and at 10 kHz, we have found in these cases that the rotation speed is likewise linearly dependent on $V^2$, but with different slopes of 5.4 rpm/ $V^2$, 4.7 rpm/ $V^2$, 4.5 rpm/ $V^2$, and 3 rpm/ $V^2$ for Au, Ni, Pt, and MWCNT respectively as shown in Fig. 3b.  The difference among the Au, Ni, Pt nanowires with the same geometrical shape scales roughly with their conductivities, since a higher conductivity facilitates a higher induced polarization, hence a higher rotation speed.  The rotation speed of MWCNT is slower and can be attributed to the less polarization of MWCNT.  It is also noted that both ferromagnetic (Ni) and non-magnetic (Au, Pt and MWCNT) entities can be rotated.

For demonstration purposes, we have made a nanowire motor by attaching the kink of a bent nanowire onto the quartz substrate through the covalent bonding between Au and thiolated substrate.  In analogy to an electric motor, the bonding between the kink of nanowire and the substrate anchors the axel of the motor as shown in Fig. 4a.  The bent nanowire serves as the rotor while the four Au electrodes serve as the stator.  After the application of a rotating AC voltage of 10 V at 20 kHz, a dust particle was being whipped, driven, and occasionally missed by the two arms of about 16 μm in length of the bent nanowire (Fig. 4b, c, d). Note that the dust particle moves only when driven by the nanowire and hence maintaining a circular motion. The relevance of such a micromotor to microfluidic devices, microstirrer, and micro electromechanical systems (MEMS) [8] is apparent.

In summary, we have demonstrated a method of rotating metallic nanowires in suspension by using AC voltages applied to multiple electrodes. Nanowire in suspension can be rotated with a specific chirality, rotation speed, and total angle of rotation. The rotation can be instantly switched on or off by the application of the AC voltages. We have used this method to determine the viscous torque on nanowires with different length. This method can also be used to explore other phenomena occurring in systems with very low Reynolds numbers, from transport of small entities, the studies of motion of microorganisms, to MEMS.

**Acknowledgements**  This work has been supported by NSF grant no. DMR00-80031.

Figure 1: **(a)** Schematics of a nanowire suspended in DI water set to rotation by quadruple electrodes, at which four phase-shifted AC voltages are simultaneously applied but with a sequential phase shift of $90^0$. **(b)** Overlapped images at 1/30 sec interval of free (right) and one-end fixed (left) rotating Au nanowires at 2.5V, 80 kHz.

Figure 2: **(a)** Rotation speed $\omega$ of free Au nanowires at 5 and 80 kHz, and Au nanowires with one end fixed at 80 kHz as a function of $V^2$. **(b)** semi-log plot of rotation speed $\omega$ versus AC frequency at 2.5 V and 5 V for free Au nanowires. **(c)** angle of rotation and **(d)** rotation speed of Au nanowires at 5 kHz and 10 V under repeated reversal of chirality.

Figure 3: **(a)** The dependence of $\omega/V^2$ and **(b)** $V^2L^3/\omega$ on the length $L$ of Au nanowires with 300 nm diameter. The curves are best-fit results to the data (see text).

Figure 4: **(a)** Snapshots of enhanced images, every 1/15 sec, of a multiwall carbon nanotube rotating at 6 V, 10 kHz. **(b)** Rotation speed $\omega$ of Au, Ni, and Pt nanowires and multiwall carbon nanotubes as a function of $V^2$.

Figure 5: **(a)** Schematics of a bent nanowire attached to the surface. Snap shots **(b)** **(c)** **(d)** of rotating bent nanowires taken every 1/30 sec and **(e)** overlapped images taken within 1.8 sec under 10V at 20 kHz illustrating a bent nanowire as a micromotor driving a dust particle.

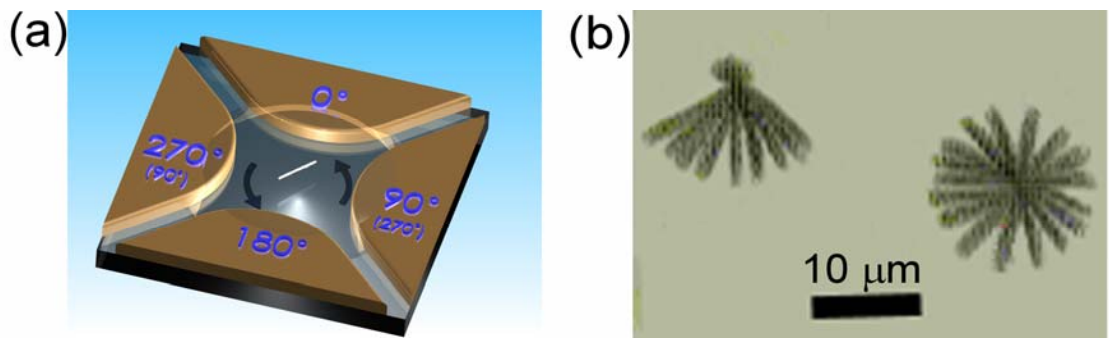

Figure 1

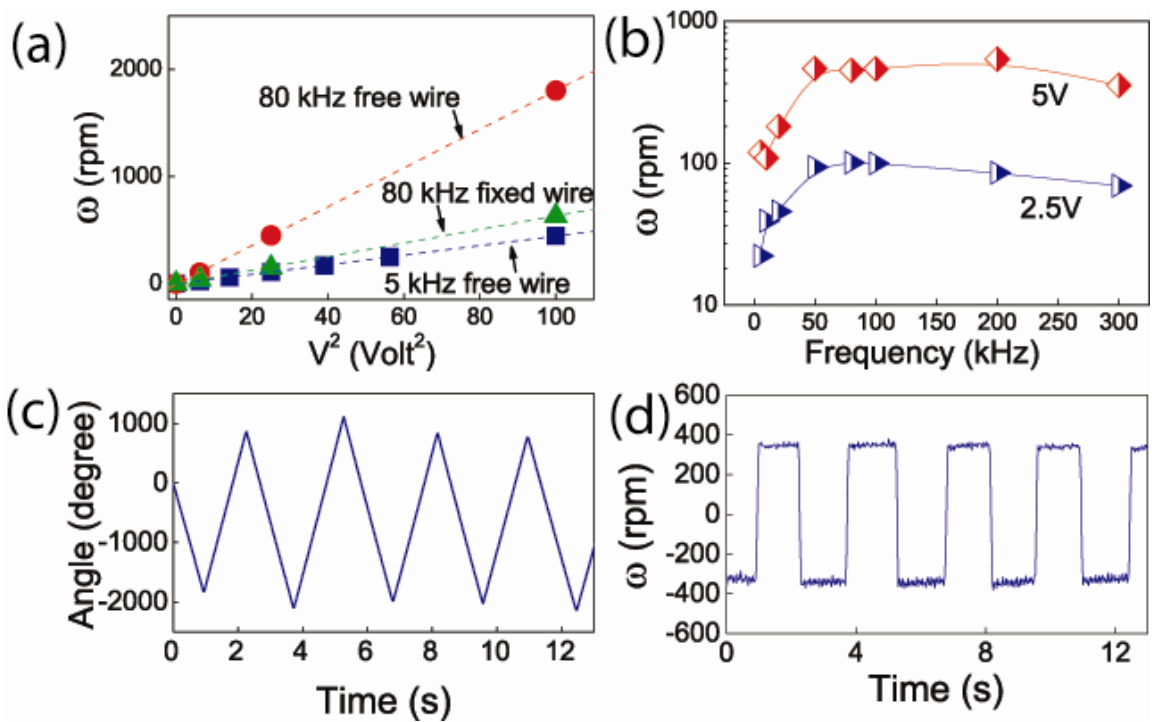

Figure 2

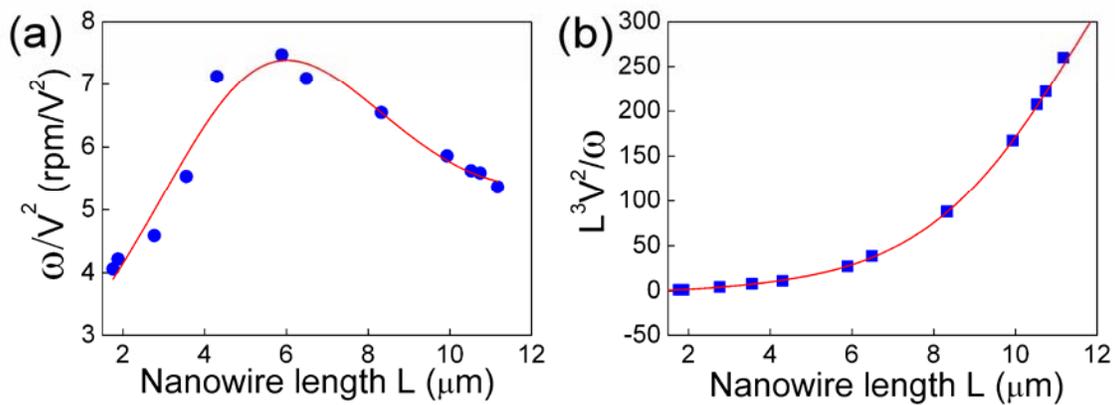

Figure 3

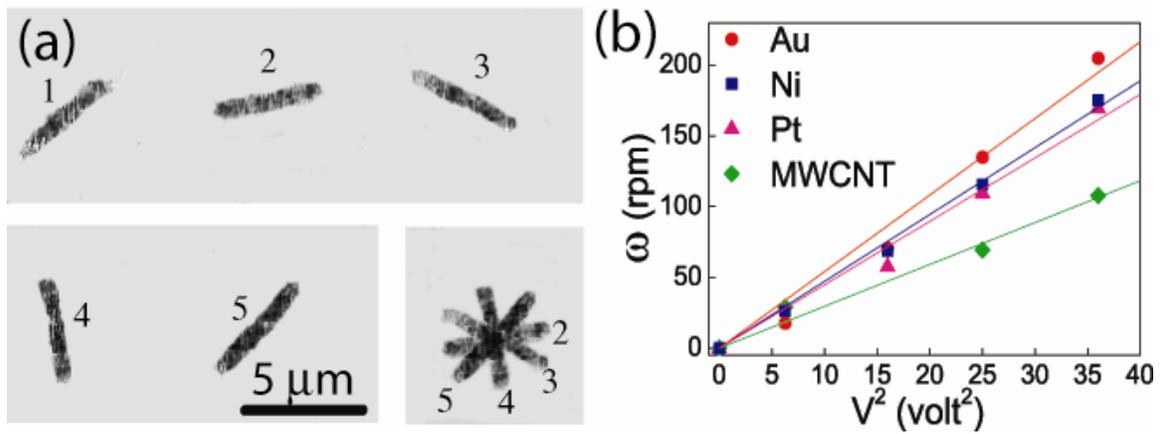

Figure 4

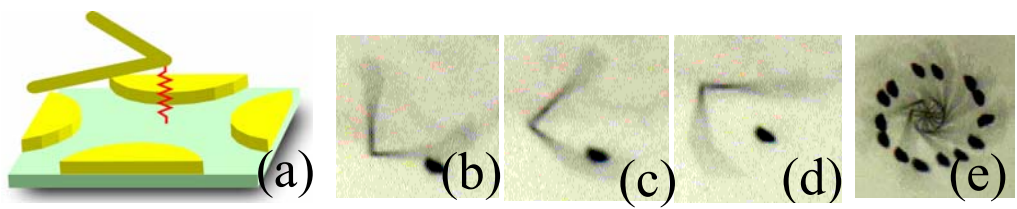

Figure 5